\documentclass[pdftex]{pasj01}

\usepackage{lineno}

\newcommand{\pasa}{PASA}
\newcommand{\actaa}{Acta Astron.}
\newcommand{\jcap}{JCAP}

\newcommand{\msun}{M_\odot}
\newcommand{\zsun}{Z_\odot}

\newcommand{\pyr}{{\rm yr}^{-1}}
\newcommand{\cgpc}{{\rm Gpc}^{-3}}
\newcommand{\mzams}{m_{\rm zams}}
\newcommand{\nhalo}{n_{\rm halo}}
\newcommand{\dthree}{\rho_{3}}
\newcommand{\mlave}{\overline{M}}
\newcommand{\mlmax}{M_{\max}}
\newcommand{\msmin}{m_{\min}}
\newcommand{\rate}{R}
\newcommand{\ratetarget}{R_{20-40}}
\newcommand{\fcemp}{f_{\rm c-rich}}
\newcommand{\msim}{{\cal M}_{\rm sim}}
\newcommand{\nsim}{{\cal N}_{\rm sim}}
\newcommand{\nsimtarget}{{\cal N}_{\rm sim, 20-40}}
\newcommand{\td}{t_{\rm d}}
\newcommand{\dtd}{\Delta \td}
\newcommand{\mrem}{m_{\rm rem}}
\newcommand{\mrapid}{m_{\rm rapid}}
\newcommand{\mc}{m_{\rm c}}
\newcommand{\mcppi}{m_{\rm c,PPI}}
\newcommand{\mcpisn}{m_{\rm c,PISN}}
\newcommand{\mcdc}{m_{\rm c,DC}}
\newcommand{\teff}{T_{\rm eff}}
\newcommand{\mbp}{m_{\rm b,p}}

\begin{document} 
\Received{}
\Accepted{}

\title{Can Population III stars be major origins of both merging binary
  black holes and extremely metal poor stars?}

\author{Ataru \textsc{Tanikawa}\altaffilmark{1}}%
\author{Gen \textsc{Chiaki}\altaffilmark{2}}
\author{Tomoya \textsc{Kinugawa}\altaffilmark{3}}
\author{Yudai \textsc{Suwa}\altaffilmark{1,4,5}}
\author{Nozomu \textsc{Tominaga}\altaffilmark{6,7,8}}

\altaffiltext{1}{Department of Earth Science and Astronomy, Graduate
  School of Arts and Sciences, The University of Tokyo, 3-8-1 Komaba,
  Meguro-ku, Tokyo 153-8902, Japan}
\altaffiltext{2}{Astronomical Institute, Graduate School of Science,
  Tohoku University, Aoba, Sendai 980-8578, Japan}
\altaffiltext{3}{Institute for Cosmic Ray Research, The University of
  Tokyo, Kashiwa, Chiba 277-8582, Japan}
\altaffiltext{4}{Department of Astrophysics and Atmospheric Sciences,
  Faculty of Science, Kyoto Sangyo University, Kyoto 603-8555, Japan}
\altaffiltext{5}{Center for Gravitational Physics, Yukawa Institute
  for Theoretical Physics, Kyoto University, Kyoto 606-8502, Japan}
\altaffiltext{6}{National Astronomical Observatory of Japan, National
  Institutes of Natural Sciences, 2-21-1 Osawa, Mitaka, Tokyo
  181-8588, Japan}
\altaffiltext{7}{Department of Physics, Faculty of Science and
Engineering, Konan University, 8-9-1 Okamoto,
Kobe, Hyogo 658-8501, Japan}
\altaffiltext{8}{Kavli Institute for the Physics and Mathematics of the
Universe (WPI), The University of Tokyo, 5-1-5 Kashiwanoha, Kashiwa, Chiba
277-8583, Japan}
  
\email{tanikawa@ea.c.u-tokyo.ac.jp}



\maketitle

\begin{abstract}

  Population (Pop) III stars, first stars, or metal-free stars are
  made of primordial gas. We have examined if they can be dominant
  origins of merging binary black holes (BHs) and extremely metal-poor
  stars. The abundance pattern of EMP stars is helpful to trace back
  the properties of Pop III stars. We have confirmed previous
  arguments that the observed BH merger rate needs Pop III star
  formation efficiency 10 times larger than theoretically predicted
  values, while the cosmic reionization history still permits such a
  high Pop III star formation efficiency. On the other hand, we have
  newly found that the elemental abundance pattern of EMP stars only
  allows the Pop III initial mass function with the minimum mass of
  $\sim 15 - 27$ $\msun$. In other words, the minimum mass must not
  deviate largely from the critical mass below and above which Pop III
  stars leave behind neutron stars and BHs, respectively. Pop III
  stars may be still a dominant origin of merging binary BHs but our
  study has reduced the allowed parameter space under a hypothesis
  that EMP stars are formed from primordial gas mixed with Pop III
  supernova ejecta.
  
\end{abstract}


\section{Introduction}

Recently, a large number of black hole (BH) mergers have been detected
by gravitational wave (GW) observations \citep{2019PhRvX...9c1040A,
  2021PhRvX..11b1053A, 2021arXiv211103634T}. Many formation scenarios
of such BH mergers have been suggested: isolated binary evolution of
Population I/II (Pop I/II) stars \citep{1998ApJ...506..780B,
  2007ApJ...662..504B,2016Natur.534..512B, 2017MNRAS.472.2422M,
  2017PASA...34...58E, 2018ApJ...866..151A, 2019MNRAS.482..870E,
  2019MNRAS.487....2M, 2020A&A...636A.104B, 2020ApJ...901L..39O,
  2021MNRAS.502.4877S, 2021A&A...649A.114G}, dynamical interactions in
globular clusters \citep{1993Natur.364..421K, 1993Natur.364..423S,
  2000ApJ...528L..17P, 2010MNRAS.402..371B, 2010MNRAS.407.1946D,
  2013MNRAS.435.1358T, 2017PASJ...69...94F, 2017MNRAS.464L..36A,
  2017MNRAS.469.4665P, 2018PhRvD..98l3005R, 2018ApJ...855..124S,
  2020MNRAS.498.4287H, 2021ApJ...915L..35K, 2021MNRAS.504.5778W,
  2020PASA...37...44A, 2021ApJ...920..128A}, star clusters
\citep{2014MNRAS.441.3703Z, 2017MNRAS.467..524B, 2017ApJ...840L..24F,
  2017PhRvD..95l4046G, 2019MNRAS.483.1233R, 2020MNRAS.495.4268K,
  2020MNRAS.498..495D, 2020ApJ...903...45K, 2021MNRAS.504..910T,
  2021ApJ...913L..29F}, and galactic centers
\citep{2009MNRAS.395.2127O, 2012ApJ...757...27A, 2016ApJ...828...77V,
  2017ApJ...835..165B, 2017ApJ...846..146P, 2017MNRAS.464..946S,
  2018ApJ...856..140H, 2018ApJ...865....2H, 2018ApJ...866...66M,
  2018MNRAS.474.5672L, 2019ApJ...876..122Y, 2019ApJ...881...20R,
  2020ApJ...899...26T, 2020MNRAS.498.4088M, 2020ApJ...891...47A,
  2021Symm...13.1678M, 2021NatAs...5..749G, 2021MNRAS.506.5451L},
secular evolution in multiple stellar systems
\citep{2014ApJ...781...45A, 2017ApJ...836...39S, 2018ApJ...863....7R,
  2019MNRAS.483.4060L, 2019MNRAS.486.4781F, 2020ApJ...898...99H,
  2020ApJ...895L..15F, 2021MNRAS.505.3844B, 2021arXiv211106388T},
hybrid channels of binary evolution and dynamical interactions
\citep{2021MNRAS.507.5224B}, and primordial BHs
\citep{2021arXiv210503349F}.

The first-generation metal-free (Population or Pop III) stars are
typically massive ($10$--$1000~M_{\odot}$) \citep{1998ApJ...508..141O,
  2002Sci...295...93A, 2004ARA&A..42...79B, 2008Sci...321..669Y,
  2011Sci...334.1250H, 2011MNRAS.413..543S, 2013ApJ...773..185S,
  2014ApJ...781...60H}, and thus leave behind BHs more efficiently
than Pop I/II stars. Thus, Pop III stars can contribute to the
formation of merging BHs \citep{2014MNRAS.442.2963K,
  2016MNRAS.460L..74H, 2017MNRAS.471.4702B, 2021ApJ...910...30T,
  2020MNRAS.495.2475L}. Moreover, several rare GW events can be
explained by Pop III stars. For example, a GW event with a $2.6$
$\msun$ compact object GW190814 \citep{2020ApJ...896L..44A} can be
explained by Pop III remnant mergers \citep{2021PTEP.2021b1E01K}.
GW190521 with a BH in the pair instability (PI) mass gap
(\cite{2020PhRvL.125j1102A, 2020ApJ...900L..13A}; but see
\cite{2020ApJ...904L..26F}) can be formed from Pop III stars
\citep{2020ApJ...903L..40L, 2021MNRAS.501L..49K, 2021MNRAS.502L..40F,
  2021MNRAS.505.2170T, 2021arXiv211010846T}. Next-generation GW
observatories, such as Cosmic Explorer \citep{2019BAAS...51g..35R} and
Einstein telescope \citep{2010CQGra..27s4002P}, may detect Pop III BHs
more massive than $100$ $\msun$ \citep{2004ApJ...608L..45B,
  2021MNRAS.505L..69H}.

However, there is a debate whether Pop III stars are a major origin of
observed BH mergers. \citet{2021MNRAS.504L..28K} have claimed that Pop
III stars are a major origin of observed BH mergers with more than 20
$\msun$ (hereafter, Pop III BH merger scenario). On the other hand,
\citet{2016MNRAS.460L..74H} (see also \cite{2021ApJ...910...30T};
\cite{2021MNRAS.501..643L}) have argued that the merger rate of Pop
III BHs should be much lower than observed. This debate comes from
adopted Pop III formation rate. The former study has supposed the
total Pop III mass in the local universe to be $\sim 10^{15}$ $\msun$
$\cgpc$, which is the upper limit constrained by the cosmic
reionization \citep{2011A&A...533A..32D, 2016MNRAS.461.2722I,
  2021ApJ...919...41I}. The latter studies have chosen it as $\sim
10^{13}$ $\msun$ $\cgpc$ obtained by theoretical studies of Pop III
star formation \citep{2016MNRAS.462.3591M, 2020MNRAS.492.4386S,
  2020ApJ...897...95V}. It is difficult to determine Pop III formation
rate, since no Pop III star has been discovered so far
\citep{2015ARA&A..53..631F, 2019MNRAS.487..486M}.

In this paper, we suggest extremely metal-poor (EMP) stars as
another approach to examine if Pop III stars may be a major origin of
observed BH mergers. EMP stars are thought as stars with ${\rm [Fe/H]}
\lesssim -3$ and considered to form in gas clouds enriched by one or
several supernovae (SNe) of Pop III stars \citep{1995ApJ...451L..49A,
  1996ApJ...471..254R, 2004A&A...416.1117C,
  2015ARA&A..53..631F}.\footnote{${\rm [X/Y]} = \log(n_{\rm X} /
  n_{\rm Y}) - \log(n_{\rm X} / n_{\rm Y})_{\odot}$, where $n_{\rm X}$
  and is the number density of an element X.}  The metal contents and
elemental abundances of EMP stars are crucial to trace back the
properties of Pop III stars. EMP stars can be divided into two types:
carbon-normal EMP stars and carbon-rich EMP stars (or carbon-enhanced
metal-poor stars, CEMP stars) \citep{2005ARA&A..43..531B}.
Carbon-normal EMP stars have elemental abundance patterns similar to
the Sun.  On the other hand, carbon-rich EMP stars have higher carbon
abundances than the Sun (${\rm [C/Fe]} > 0.7$)
\citep{2007ApJ...655..492A}. The abundance ratio of carbon-rich stars
to carbon-normal stars increases with [Fe/H] decreasing
\citep{2005ARA&A..43..531B, 2013ApJ...762...26Y}.

The presence of the two types can be interpreted as follows.
Elemental abundances of carbon-normal EMP stars can best fit with the
models of {\it successful} core-collapse supernovae (CCSNe).  To
explain the origin of carbon-rich stars, \citet{2003Natur.422..871U}
have proposed a so-called faint SN model: inner layers of stellar core
falls back into central compact remnants, and only outer layers
containing light elements, such as carbon, are ejected.  However, in
their model, mass cut is a free parameter to reproduce the elemental
abundance of carbon-rich stars.  \citet{2012ApJ...749...91F} have
independently found that a considerable fraction of ejecta falls back
into a proto-neutron stars in a window of progenitor mass
$20$--$30~M_{\odot}$.  Such {\it failed} supernovae (FSNe) leave
behind a BH and should eject only carbon-rich outer layers. There is a
firm evidence that a CEMP star is formed from the mixture of
primordial gas and the ejecta of FSNe
\citep{2013ApJ...773...33I}. These studies motivate us to define CCSNe
and FSNe as the progenitors of carbon-normal and carbon-rich EMP
stars, respectively. If primordial gas in dark matter minihalos
(hereafter, minihalos) is enriched by carbon-normal and carbon-rich
materials, the minihalos form carbon-normal and carbon-rich EMP stars,
respectively. Note that there are many other scenarios of carbon-rich
star formations: abundant production of carbon in supermassive rapidly
rotating stars \citep{2001ApJ...550..372F, 2021MNRAS.506.5247L},
enrichment of a Pop III star from asymptotic giant branch winds
\citep{2004ApJ...611..476S, 2010A&A...522L...6C}, and interstellar
medium accretion of gas decoupled from dust
\citep{2015MNRAS.453.2771J}.

The purpose of this paper is to assess the Pop III BH merger scenario,
under a hypothesis that EMP stars consist of primordial gas and Pop
III supernova ejecta. Here, we remark the importance of this
purpose. Although many formation scenarios of BH mergers have been
proposed as described above, they have not yet been critically
verified, and none of them has been rejected. Moreover, several
studies have suggested that binary BHs can have multiple origins
\citep{2021MNRAS.507.5224B, 2021ApJ...910..152Z, 2021ApJ...913L...5N,
  2021JCAP...03..068H}. As the first step, we need to narrow down
promising formation scenarios of BH mergers, and scrutinize these
formation scenarios one by one.  In particular, we focus on Pop III
stars, since they can be the origin of binary BHs with $\sim 30$
$M_\odot$ \citep{2014MNRAS.442.2963K, 2021MNRAS.504L..28K}, or with PI
mass gap BHs \citep{2021MNRAS.505.2170T}.  Among the scenarios of Pop
III star origins, we pay attention to the Pop III BH merger scenario
suggested by \cite{2014MNRAS.442.2963K} and
\cite{2021MNRAS.504L..28K}.  Since this scenario assumes the largest
Pop III star mass in total among all the BH scenarios requiring Pop
III stars, it would be easiest to get constrained. We also attribute
CCSNe and FSNe to the formation of EMP stars, although there are
several scenarios to explain the formation of EMP stars.  This is
because this EMP formation scenario can explain the abundance pattern
of EMP stars \citep{2013ApJ...773...33I} as described above.

The structure of this paper is as follows. In section
\ref{sec:Method}, we present our method to obtain the merger rate of
Pop III BHs, and the abundance ratio of carbon-rich EMP stars to
carbon-normal EMP stars. In sections \ref{sec:Results} and
\ref{sec:SummaryAndDiscussion}, we show our results, and summarize
this paper, respectively.

\section{Method}
\label{sec:Method}

In this section, we present a method to obtain the merger rate of Pop
III BHs, and the abundance ratio of carbon-rich EMP stars to
carbon-normal EMP stars. In section
\ref{sec:PopIIIStarFormationModel}, we describe our Pop III star
formation model. In section \ref{sec:BH-BHFormationModel}, we review
our binary population synthesis model to derive the merger rate of Pop
III BHs. In section \ref{sec:EMPStarFormationModel}, we show our EMP
star formation model based on the results of binary population
synthesis calculations. In section \ref{sec:NecessaryConditions}, we
describe the necessary conditions satisfying GW and EMP observations.

\subsection{Pop III star formation model}
\label{sec:PopIIIStarFormationModel}

We assume that Pop III stars are born in minihalos. The minihalos are
supposed to have $\sim 10^6 \msun$ and be formed at the high-redshift
universe, $z \gtrsim 5$ \citep{2016MNRAS.460L..74H,
  2020MNRAS.492.4386S, 2020ApJ...897...95V}. We adopt the number
density of the minihalos in the local universe ($\nhalo$), such that
$\nhalo \sim 10^{11}$ $\cgpc$. We obtain this number density from the
results of \citet{2020MNRAS.492.4386S}; their figure 4 shows that the
total mass of Pop III stars formed in each 1 cubic Gpc is $\sim 3
\times 10^{13}$ $\msun$ $\cgpc$, and their figure 8 indicates that the
total mass of Pop III stars in each minihalo ($\mlave$) is $\sim 300$
$\msun$ on average. We basically refer to the Pop III formation model
of \citet{2020MNRAS.492.4386S}. Although we adopt the model of
\citet{2020MNRAS.492.4386S}, we note that many research groups have
studied the Pop III formation in the universe
\citep{2014ApJ...792...32S, 2015MNRAS.448..568H, 2016ApJ...826....9I,
  2016MNRAS.461.2722I, 2016MNRAS.462.3591M, 2020ApJ...897...95V}.

We construct the mass function of the total mass of Pop III stars in
each minihalo ($M$), expressed as
\begin{eqnarray}
f_1(M) \propto M^{-1} \; (0.1 \mlmax < M < \mlmax).
\end{eqnarray}
We get this mass function, simplifying that of
\citet{2020MNRAS.492.4386S} (see their figure 8) for ease of
handling. Their mass function can be interpreted as logarithmic flat
over a single digit (from $\sim 50 \msun$ to $\sim 500 \msun$). We
ignore that a small fraction of minihalos form $\lesssim 50 \msun$ Pop
III stars in total. We treat the maximum mass of Pop III stars in
total ($\mlmax$) as a free parameter, although $\mlmax \sim 500 \msun$
in \citet{2020MNRAS.492.4386S}. A reason for this treatment is that
Pop III star formation models in each minihalo still have
uncertainties in baryon physics, such as magnetic field, turbulence,
and so on \citep{2011ApJ...731...62F, 2012ApJ...745..154T,
  2021MNRAS.505.4197S, 2021ApJ...915..107H}.  Moreover, we practically
need Pop III stars enough to explain the observed BH merger rate.

We suppose that all the Pop III stars are born at the same time at the
high-redshift of $\gtrsim 5$. We set the binary fraction of Pop III
stars to 1. This is because Pop III stars are usually born in multiple
stellar systems \citep{2009Sci...325..601T, 2011Sci...331.1040C,
  2011ApJ...737...75G, 2014ApJ...792...32S, 2020ApJ...892L..14S}.

The initial mass function (IMF) of primary stars is expressed as
\begin{eqnarray}
  f_2(m_1) \propto m_1^{-1} \; (\msmin < m_1 < 150 \msun).
\end{eqnarray}
The distribution is logarithmically flat, which is consistent with
\citet{2014ApJ...792...32S} and \citet{2014ApJ...781...60H}. Although
a part of Pop III stars may have more than $150$ $\msun$ (but see
\cite{2020ApJ...897...58T}), we do not take into account them
conservatively. Note that such Pop III stars form more massive than
$100$ $\msun$ BHs, and thus do not contribute to the observed BH
merger rate. We regard the minimum mass of Pop III stars as a free
parameter. This parameter has large effects on the elemental abundance
pattern of EMP stars.

Binary parameters we adopt are the same as those proposed by
\citet{2012Sci...337..444S}. The mass ratio of companion stars to
primary stars ($q$) can be expressed as
\begin{eqnarray}
  f_3(q) \propto q^{-0.1} (0.1 < q < 1).
\end{eqnarray}
Note that the minimum mass of the companion stars is also
$\msmin$. The binary period ($P$) and eccentricity ($e$) distribution
are, respectively, given by
\begin{eqnarray}
  f_4(\log P) \propto \left( \log P \right)^{-0.55} \; (0.15 <
  \log(P/{\rm day}) < 5.5),
\end{eqnarray}
and
\begin{eqnarray}
  f_5(e) \propto e^{-0.42} \; (0 < e < 1).
\end{eqnarray}
Note that the minimum and maximum periods are the same as those of
\cite{2015ApJ...814...58D}.

\subsection{Merging BH formation model}
\label{sec:BH-BHFormationModel}

We generate $10^6$ binary stars with the primary IMF and binary
parameters described in section \ref{sec:PopIIIStarFormationModel}. We
follow the evolution of the binary stars, using binary population
synthesis technique. We calculate the BH merger rate density in the
current universe ($\rate$) as
\begin{eqnarray}
  \rate = \frac{\dthree}{\msim} \frac{\nsim (\td)}{\dtd},
\end{eqnarray}
where $\dthree$ is the total mass of Pop III stars formed in each unit
volume, $\msim$ is the total mass of the simulated binary stars, $\td$
is the delay time of BH mergers from Pop III star formation, and
$\nsim (\td)$ is the total number of BH mergers with the delay time
from $\td$ to $\td+\dtd$ in the simulations. We set $\td=10$~Gyr and
$\dtd=5$~Gyr, since Pop III stars are born in the early universe. As
described later, we focus on BH mergers with the heavier mass of
$20-40$ $\msun$. The BH merger rate can be calculated as
\begin{eqnarray}
  \ratetarget = \frac{\dthree}{\msim} \frac{\nsimtarget (\td)}{\dtd},
\end{eqnarray}
where $\nsimtarget (\td)$ is the total number of BH mergers with the
heavier mass of $20-40$ $\msun$, and with the delay time from $\td$ to
$\td+\dtd$ in the simulations.

Hereafter, we describe our binary population synthesis technique. We
use a binary population synthesis code {\tt BSE} with extensions to
EMP stars \citep{2020MNRAS.495.4170T, 2021ApJ...910...30T}. Pop III
stars evolve along with the fitting formulae of stars with stellar
metallicity $Z=10^{-8}$ $\zsun$. The evolution of $Z=10^{-8}$ $\zsun$
stars is the same as Pop III star evolution because of the low
metallicity. We do not consider stellar wind mass loss because of zero
metallicity. Our supernova model is based on the rapid model of
\citet{2012ApJ...749...91F} with modifications of PI. The PI model is
given by
\begin{eqnarray}
  \mrem = \displaystyle \left\{
  \begin{array}{ll}
    \mrapid  & (\mc \le \mcppi) \\
    \mcppi   & (\mcppi  < \mc \le \mcpisn) \\
    0        & (\mcpisn < \mc \le \mcdc) \\
    \mrapid  & (\mc > \mcdc)
  \end{array}
  \right., \label{eq:PImodel}
\end{eqnarray}
where $\mrem$ is the remnant mass, $\mrapid$ is the remnant mass
without the PI model, and $\mc$ is the helium core mass. In the
ascending order of the helium core mass, stars experience supernovae
in the framework of the rapid model, pulsational PI
\citep{2002ApJ...567..532H, 2007Natur.450..390W, 2016MNRAS.457..351Y,
  2017ApJ...836..244W, 2019ApJ...887...72L}, PI supernovae
\citep{1967PhRvL..18..379B, 1968Ap&SS...2...96F, 1984ApJ...280..825B,
  1986A&A...167..274E, 2001ApJ...550..372F, 2002ApJ...567..532H,
  2002ApJ...565..385U, 2011ApJ...734..102K, 2012A&A...542A.113Y}, and
collapse to a BH together with possible association with gamma-ray
bursts \citep{2001ApJ...550..372F, 2006ApJ...645..519N,
  2007ApJ...665L..43S, 2007PASJ...59..771S, 2009ApJ...690..913S,
  2011ApJ...726..107S, 2012ApJ...752...32W, 2012ApJ...759..128N,
  2019ApJ...870...98U, 2019PhRvD..99d1302U}. We adopt $(\mcppi,
\mcpisn, \mcdc) = (45 \msun, 65 \msun, 135 \msun)$, which is similar
to the model of \citet{2016A&A...594A..97B}.  We do not account for BH
natal kicks. BH natal kicks have small effects on merging binary BHs,
since their binary motions are much faster than their natal kicks
\citep{2021ApJ...910...30T}.

As described above, stars experience supernovae in the framework of
the rapid model if $\mc \le \mcppi$, or if $\mzams \lesssim 80$
$\msun$, where $\mzams$ is zero-age main sequence (ZAMS) mass. The
supernovae can be divided into three types by $\mzams$. For $\mzams
\lesssim 20$ $\msun$, stars leave behind neutron stars with supernova
ejecta. We regard that these stars cause CCSNe.  For $20 \lesssim
\mzams \lesssim 30$ $\msun$, stars leave behind BHs with supernova
ejecta. We regard that these stars cause FSNe. For $30 \lesssim
\mzams/\msun \lesssim 80$, stars leave behind BHs without supernova
ejecta, which can be interpreted as direct collapse.

Our binary evolution model is based on the {\tt BSE} code
\citep{2002MNRAS.329..897H}. We describe the difference between the
original {\tt BSE} and our models.  In the original {\tt BSE} model,
core helium burning and shell helium burning stars are assumed to have
radiative and convective envelopes, respectively. On the other hand,
in our model, stars with $\log(\teff/K) \ge 3.65$ and $< 3.65$ are
assumed to have radiative and convective envelopes, respectively,
where $\teff$ is the effective temperature of a star.  The original
{\tt BSE} and our models have different formulae of tidal interaction
for stars with radiative envelopes. The original {\tt BSE} model
adopts the formulae of \citet{1975A&A....41..329Z}, while our model
adopts the formulae of \citet{2021MNRAS.504L..28K}, based on
\citet{2010ApJ...725..940Y} and \citet{2018A&A...616A..28Q}. The mass
transfer rate in Roche-lobe overflow is calculated by the formulae of
\citet{2021MNRAS.504L..28K}, based on \citet{1972AcA....22...73P},
\citet{1978A&A....62..317S}, \citet{1987MNRAS.229..383E}, and
\citet{2017MNRAS.468.5020I}. The common-envelope evolution is modeled
as $\alpha$ formalism \citep{1984ApJ...277..355W}. We adopt
$\alpha_{\rm CE}=1$ and calculate $\lambda_{\rm CE}$ from
\citet{2014A&A...563A..83C}. We omit magnetic braking, since the
magnetic field of first stars is expected to weaker than that of Pop
I/II stars \citep{1989ApJ...342..650P,2011ApJ...741...93D} and the
dominant component of the magnetic field is tangled according to
\citet{2020MNRAS.497..336S}.

\subsection{EMP star formation model}
\label{sec:EMPStarFormationModel}

We suppose that EMP stars are born in minihalos
\citep{2020MNRAS.497.3149C} and made of primordial gas enriched by Pop
III CCSNe and FSNe. We focus only on whether EMP stars are carbon-rich
or carbon-normal. Thus, we take into account chemical elements of
hydrogen, carbon, and iron.

Each minihalo has the following elemental abundance after all the Pop
III stars in the minihalo finish their evolution. Each minihalo has
$10^5 \msun$ hydrogen. When a Pop III star causes a CCSN, it ejects
$0.2 \msun $ carbon and $0.07 \msun$ iron
\citep{2013ARA&A..51..457N}. When a Pop III star causes a FSN, it
ejects $0.989 \msun$ carbon and $1.47 \times 10^{-5} \msun$ iron
\citep{2014ApJ...794..100M}.  When a Pop III star experiences
pulsational PI, it ejects no carbon or iron, since its carbon core is
intact \citep{2020ApJ...905L..21U}. When a Pop III star experiences a
PI supernova, it disrupts the minihalo because of its large explosion
energy, and thus the minihalo cannot form EMP stars
\citep{2007ApJ...670....1G, 2014ApJ...791..116C, 2018MNRAS.475.4378C}.

We regard that the number ratio of carbon-rich EMP stars to
carbon-normal EMP stars is equal to that of carbon-rich minihalos to
carbon-normal minihalos. In other words, we assume that chemical
elements are completely mixed in each minihalo, and that all the
minihalos form EMP stars with the same number and IMF. We adopt these
conditions for simplicity, although IMF of EMP stars can depend on
metallicity \citep{2000ApJ...534..809O, 2003Natur.422..869S,
  2013ApJ...766..103D, 2016MNRAS.463.2781C, 2021MNRAS.508.4175C}.

\subsection{Necessary conditions from Pop III, GW and EMP observations}
\label{sec:NecessaryConditions}

In this section, we describe three necessary conditions where Pop III
stars are major origins of merging BHs whose masses are 20--40
$\msun$, and EMP stars are formed from primordial gas mixed with Pop
III supernova ejecta. The three necessary conditions are as follows.
\begin{enumerate}
\item $\dthree \le 10^{15}$ $\msun$ $\cgpc$
\item $3 \le \ratetarget/(\pyr \cgpc) \le 30$
\item $0.01 \le \fcemp \le 1 $
\end{enumerate}
We explain these conditions in detail later.

We set the first necessary condition to impose an upper limit on
$\dthree$, the total mass of Pop III stars formed in each 1 cubic Gpc
until now. The upper limit is $10^{15}$ $\msun$ $\cgpc$. More than the
upper limit cannot be permitted by reionization constraints
\citep{2016MNRAS.461.2722I, 2021ApJ...919...41I}.

We need the second necessary condition to regard Pop III BHs as a
major origin of observed BH mergers. \citet{2021MNRAS.504L..28K} have
argued that Pop III BHs can form all the observed BH mergers with the
primary masses of $> 20 \msun$. We do not take into account BH mergers
with more than $40$ $\msun$ for the second necessary condition. There
are two reasons. First, we can say that the Pop III BH merger scenario
should be successful, if the second necessary condition is
satisfied. Second, a BH merger rate with more than $40$ $\msun$
strongly depends on uncertainties of the PI model
\citep{2020ApJ...902L..36F, 2020ApJ...905L..15B,
  2021MNRAS.501.4514C}. The second necessary condition may be strict
more than necessary. As a reference, we investigate the case where
this condition is relaxed to $\ratetarget = 1-100$ $\pyr$ $\cgpc$.

We adopt the third necessary condition under which the number ratio of
carbon-rich EMP stars to carbon-normal EMP stars, $\fcemp$, is
consistent with EMP observations. We define EMP stars as those with
${\rm [Fe/H]} \le -2.5$.  Moreover, we define carbon-rich and
carbon-normal EMP stars as those with [C/Fe] $>2$ and $<2$,
respectively, following \citet{2017MNRAS.472L.115C} (see also
\cite{2016ApJ...833...20Y}). According to the SAGA database
\citep{2008PASJ...60.1159S}, $\fcemp \sim 0.05$. We do not impose a
strict condition on $\fcemp$. Rather, we allow $\fcemp$ to have wide
range of values, such that $\fcemp$ ranges over 2 digits.  This is
because we obtain $\fcemp \sim 0.05$ directly from the SAGA database,
ignoring selection bias. If we take into account the completeness of
EMP star observations, we may take the necessary condition more
strictly. Eventually, we can narrow the allowed region. However, this
is beyond of the scope of this paper.

\section{Results}
\label{sec:Results}

\begin{figure*}
 \begin{center}
  \includegraphics[width=16cm, bb=0 0 432 216]{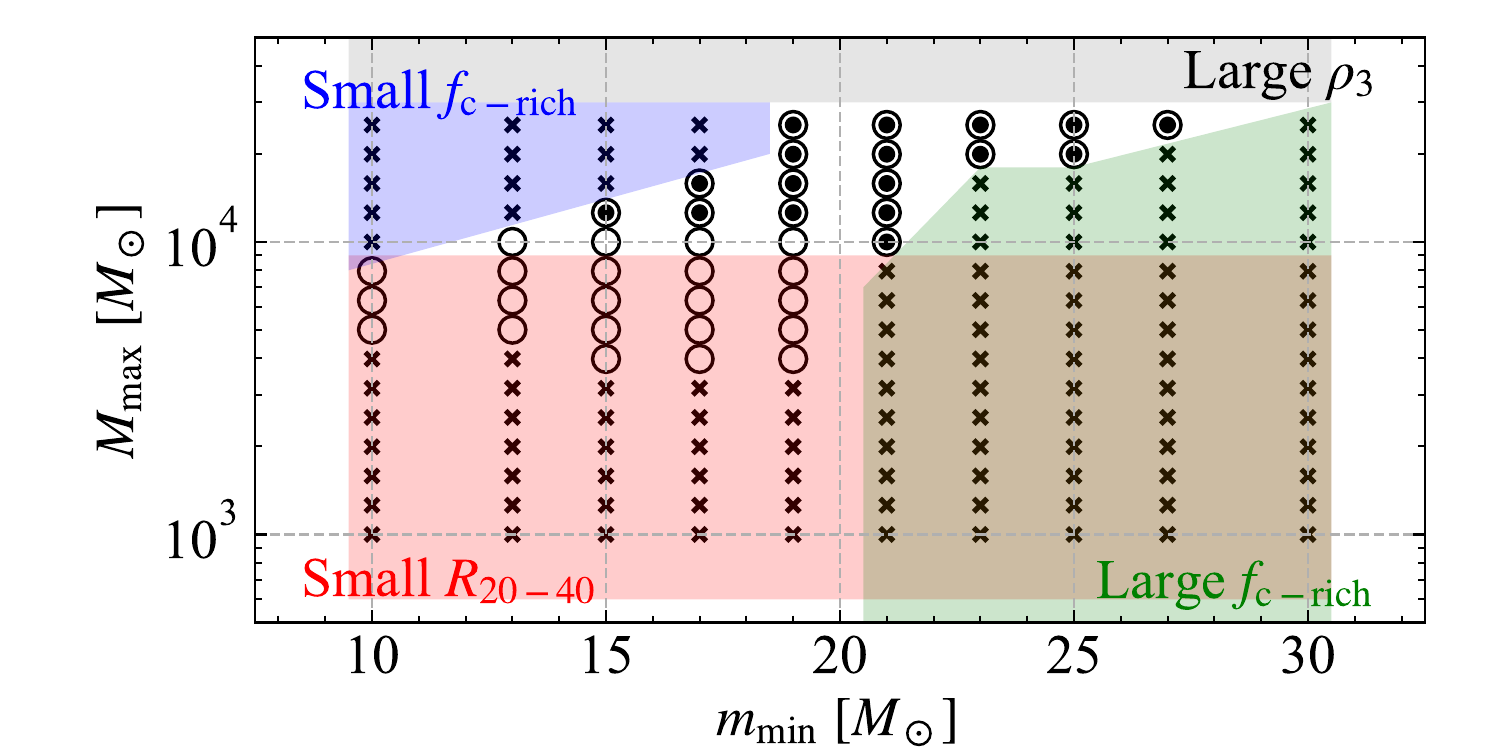}
 \end{center}
\caption{Combinations of $\msmin$ and $\mlmax$ which satisfy the three
  necessary conditions described in section
  \ref{sec:NecessaryConditions}. $\msmin$ and $\mlmax$ are the minimum
  mass of the primary stellar IMF, the maximum muss for the mass
  function of the total mass of Pop III stars in each halo,
  respectively. The combinations indicated by open circles with dots
  satisfy all the necessary conditions. Those indicated by open
  circles do so, when the second necessary is relaxed to $1 <
  \ratetarget/(\pyr \cgpc) < 10^2$. Those indicated by crosses do
  not. Shaded regions indicate which necessary condition the
  combinations do not satisfy. In the red, blue, green, and gray
  regions, $\ratetarget$ is smaller, $\fcemp$ is smaller, $\fcemp$ is
  larger, and $\dthree$ is larger than the corresponding necessary
  conditions, respectively.} \label{fig:figDiagram}
\end{figure*}

Figure \ref{fig:figDiagram} shows which ($\msmin$, $\mlmax$) satisfies
the three necessary conditions described in section
\ref{sec:NecessaryConditions}. We first focus on $\mlmax$. We need
$\mlmax \lesssim 3 \times 10^4$ $\msun$ in order to satisfy the first
necessary condition. We also need $\mlmax \gtrsim 10^4$
$\msun$. Otherwise, the second necessary condition cannot be
satisfied, since the Pop III BH merger rate density is too small:
$\ratetarget < 3$ $\pyr$ $\cgpc$. This $\mlmax$ is much larger than
$\mlmax \sim 600$ $\msun$ obtained by numerical simulations of
\citet{2020MNRAS.492.4386S}. Similar things have been already pointed
out by \citet{2016MNRAS.460L..74H}. Nevertheless, we have to note that
the first necessary condition is not still violated even when the
second necessary condition is satisfied.

We next focus on $\msmin$.  We need $15 \lesssim \msmin/\msun \lesssim
27$ in order to satisfy the third necessary condition. The allowed
$\mlmax$ becomes narrower when $\msmin$ deviates from $\sim 20 \msun$,
the boundary mass between CCSNe and FSNe. If $\msmin$ is too small,
$\fcemp < 0.01$, since minihalos have a large number of CCSNe, get
iron elements, and cannot form carbon-rich EMP stars.  If $\msmin$ is
too large, $\fcemp > 1$, since minihalos have a small number of CCSNe,
get little iron element, and cannot form carbon-normal EMP stars.

If we relax the second necessary condition from $3 \le
\ratetarget/(\pyr \cgpc) \le 30$ to $1 \le \ratetarget/(\pyr \cgpc)
\le 100$, the allowed region becomes wider not only in the direction
of $\mlmax$, but also in the direction of $\msmin$. Especially,
$\msmin = 10$ $\msun$ can be allowed. Each minihalo forms a smaller
amount of Pop III stars in the latter case than in the former
case. Each minihalo has CCSNe at a smaller probability, and gets a
smaller amount of iron elements. Thus, some of minihalos can form
carbon-rich EMP stars.

\begin{table}
    \caption{$(\msmin, \mlmax)$ chosen for Figures
      \ref{fig:figExample1}, \ref{fig:figExample2}, and
      \ref{fig:figExample3}, where $\msmin$ and $\mlmax$ are the
      minimum mass of the primary stellar IMF and the maximum mass for
      the mass function of the total mass of Pop III stars in each
      halo, respectively.}
    \label{tab:msmin_mlmax}
    \centering
    \begin{tabular}{lll}
    \hline
        Figure & $\msmin$ & $\mlmax$ \\
    \hline
        Figure \ref{fig:figExample1} & $10$ $\msun$ & $600$   $\msun$ \\
        Figure \ref{fig:figExample2} & $10$ $\msun$ & $15000$ $\msun$ \\
        Figure \ref{fig:figExample3} & $17$ $\msun$ & $12000$ $\msun$ \\
    \hline
    \end{tabular}
\end{table}

\begin{figure*}
 \begin{center}
  \includegraphics[width=13cm, bb=0 0 720 360]{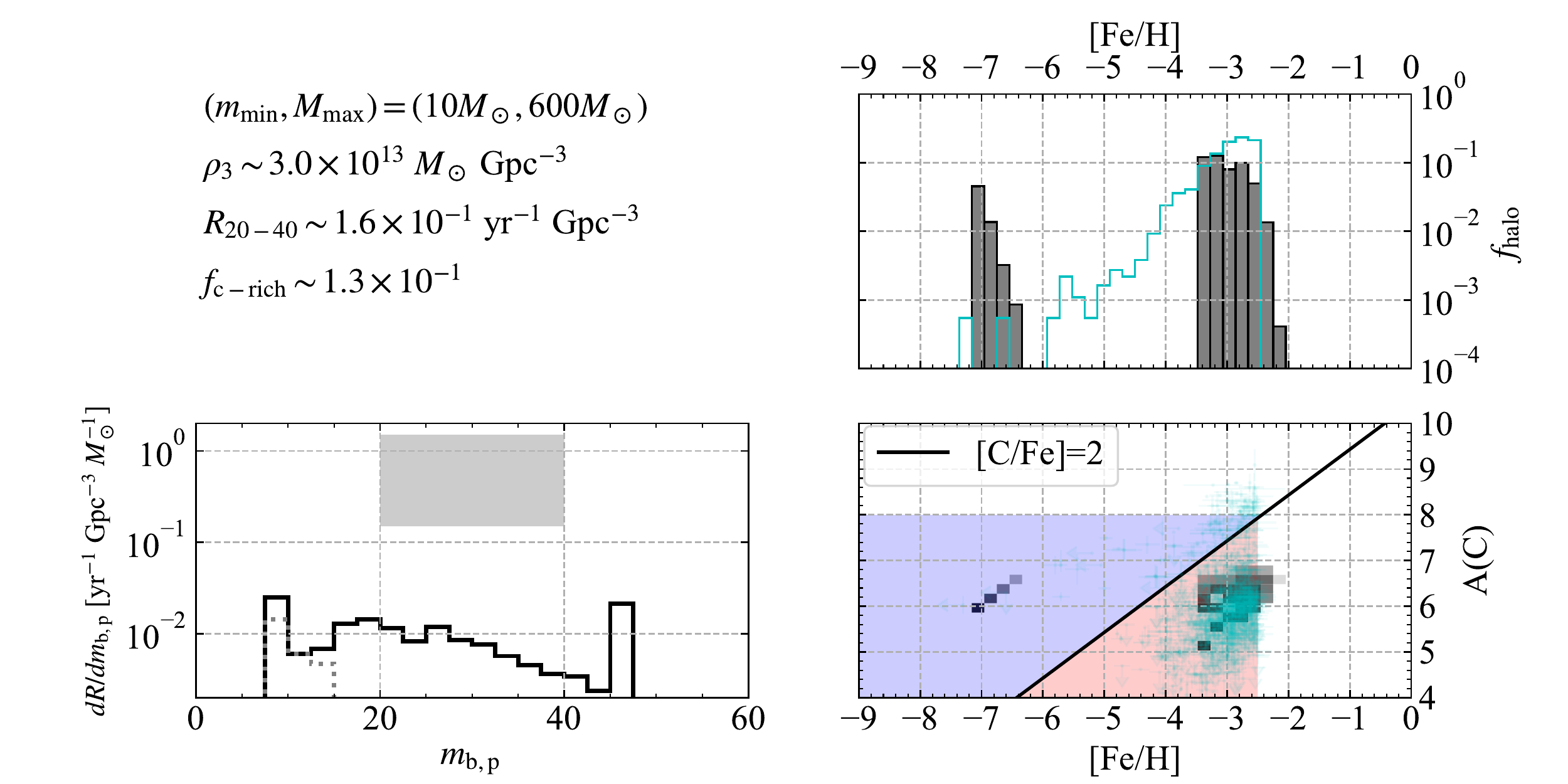} 
 \end{center}
\caption{(Top left) Input parameters of $(\msmin,
  \mlmax)=(10\msun,600\msun)$, and resulting values of $\dthree$,
  $\ratetarget$, and $\fcemp$. (Bottom left) BH merger rate density
  differentiated by the heavier BH mass ($\mbp$) in the local
  universe. Gray-shaded regions in the left panels show $3 <
  R_{20-40}/(\pyr \cgpc) < 30$ if the $dR/d\mbp$ distribution is
  assumed to be flat in this range. The dotted gray histograms
  indicate BH mergers either of which are formed through FSNe. FSNe
  form only light BHs. There are two local peaks at $m_1 \sim 10$ and
  $45$ $\msun$. The lower-mass peak is formed through the rapid model
  (\cite{2012ApJ...749...91F}) in which relatively low-mass
  progenitors ($20$--$30$ $M_\odot$ ZAMS stars) preferentially leave
  behind $5$--$10$ $M_\odot$ BHs. The higher-mass peak is formed by
  pulsational PI (see Eq. (\ref{eq:PImodel})). (Top right) [Fe/H]
  distribution of minihalos after Pop III star evolution. The
  abundance of EMP stars is indicated by the cyan histograms
  normalized to unity. These data are retrieved from the SAGA database
  (\cite{2008PASJ...60.1159S, 2011MNRAS.412..843S,
    2013MNRAS.436.1362Y, 2017PASJ...69...76S}). Our model do not
  obtain $-6 \lesssim$ [Fe/H] $\lesssim -4$, since we do not take into
  account variation of FSNe unlike \cite{2014ApJ...785...98T}.
  (Bottom right) Carbon ($A(C)$) and [Fe/H] distribution of minihalos
  after Pop III star evolution. Blue and red-shaded regions in the
  right panels indicate the abundance of carbon-rich and carbon-normal
  EMP stars, respectively. The cyan points with error bars indicate
  EMP stars.} \label{fig:figExample1}
\end{figure*}

In order to complement the above explanations, we demonstrate how to
satisfy the three necessary conditions. Table \ref{tab:msmin_mlmax}
shows our choice of $(\msmin, \mlmax)$ in advance. As the starting
point, we choose $(\msmin, \mlmax) = (10\msun, 600\msun)$. The values
of $\msmin$ and $\mlmax$ are consistent with
\citet{2014ApJ...792...32S} and \citet{2020MNRAS.492.4386S},
respectively. Thus, these values are based on recent numerical
simulations for Pop III star formations. The results are shown in
Figure \ref{fig:figExample1}. In this case, the first and third
conditions are satisfied. However, the BH merger rate is $\ratetarget
\sim 1.6 \times 10^{-1}$ yr$^{-1}$ $\cgpc$, which is much smaller than
the second necessary condition. Thus, we need to change $(\msmin,
\mlmax)$ to increase $\ratetarget$.

\begin{figure*}
 \begin{center}
  \includegraphics[width=13cm, bb=0 0 720 360]{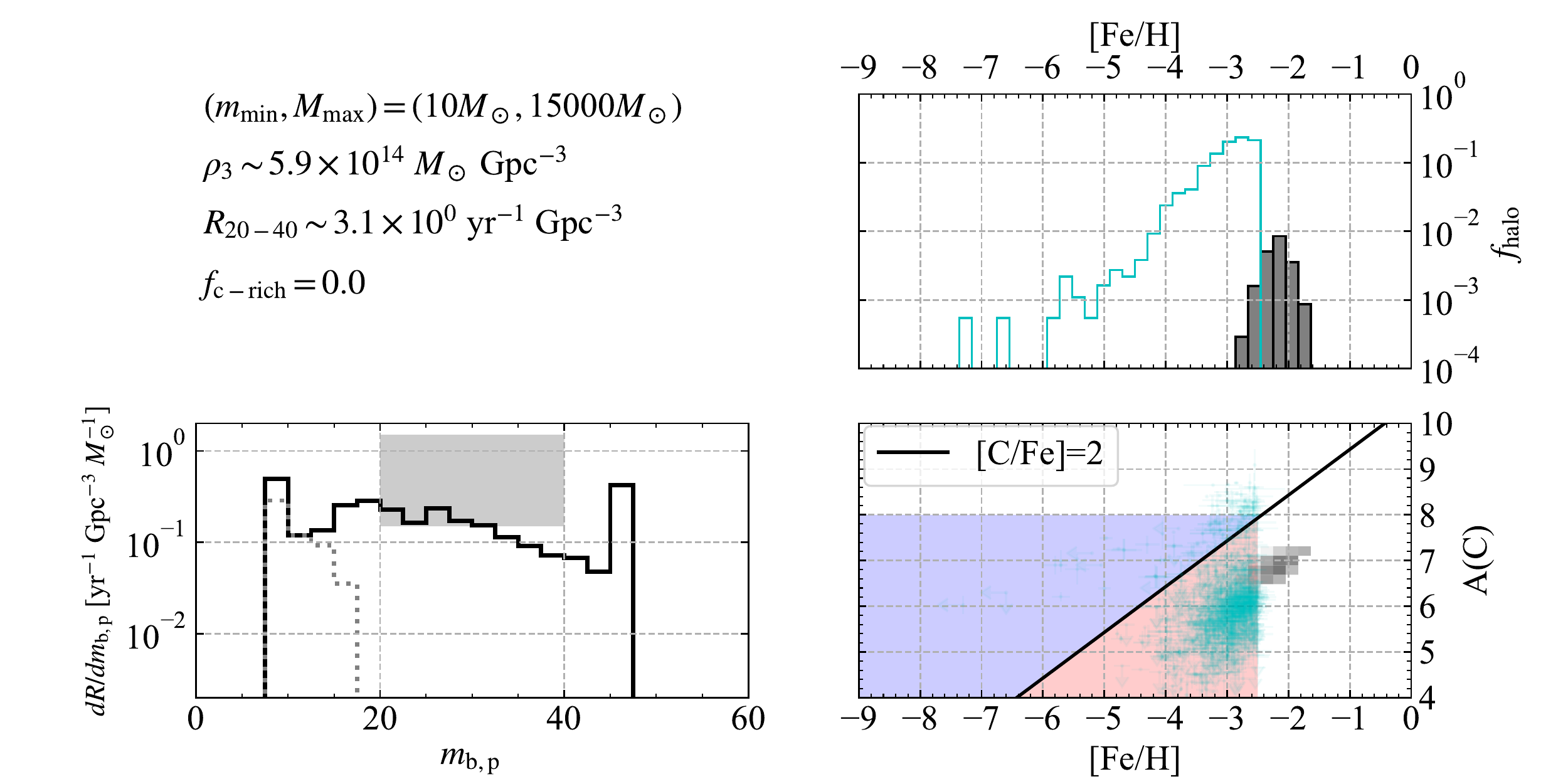} 
 \end{center}
\caption{The same as Figure \ref{fig:figExample1} except for input
  parameters of $(\msmin, \mlmax)=(10\msun,15000\msun)$. Two local
  peaks at $m_1 \sim 10$ and $45$ $\msun$ in the bottom left panel are
  formed through the same mechanisms as in
  Figure~\ref{fig:figExample1}.} \label{fig:figExample2}
\end{figure*}

In order to increase $\ratetarget$, we set $(\msmin, \mlmax) =
(10\msun, 15000\msun)$. Figure \ref{fig:figExample2} shows the results
for this case. The Pop III star formation rate is $\dthree \sim 5.9
\times 10^{14}$ $\msun$ $\cgpc$. This is much larger than that of
\citet{2020MNRAS.492.4386S}, however does not still violate the first
necessary condition. The large $\mlmax$ increases $\ratetarget$ to
$\sim 3.1$ $\pyr$ $\cgpc$, consistent with the second necessary
condition. On the other hand, $\fcemp \ll 0.01$. This is much smaller
than the third necessary condition. The reason why $\fcemp$ decreases
from the case of $(\msmin, \mlmax) = (10\msun, 600\msun)$ to
$(10\msun, 15000\msun)$ is as follows. In the case of $(\msmin,
\mlmax) = (10\msun, 15000\msun)$, all the minihalos have a large
number of Pop III stars with $\mzams \lesssim 20 \msun$. They have
many CCSNe, and are sufficiently polluted by iron. Thus, they form
only carbon-normal EMP stars after that.

\begin{figure*}
 \begin{center}
  \includegraphics[width=13cm, bb=0 0 720 360]{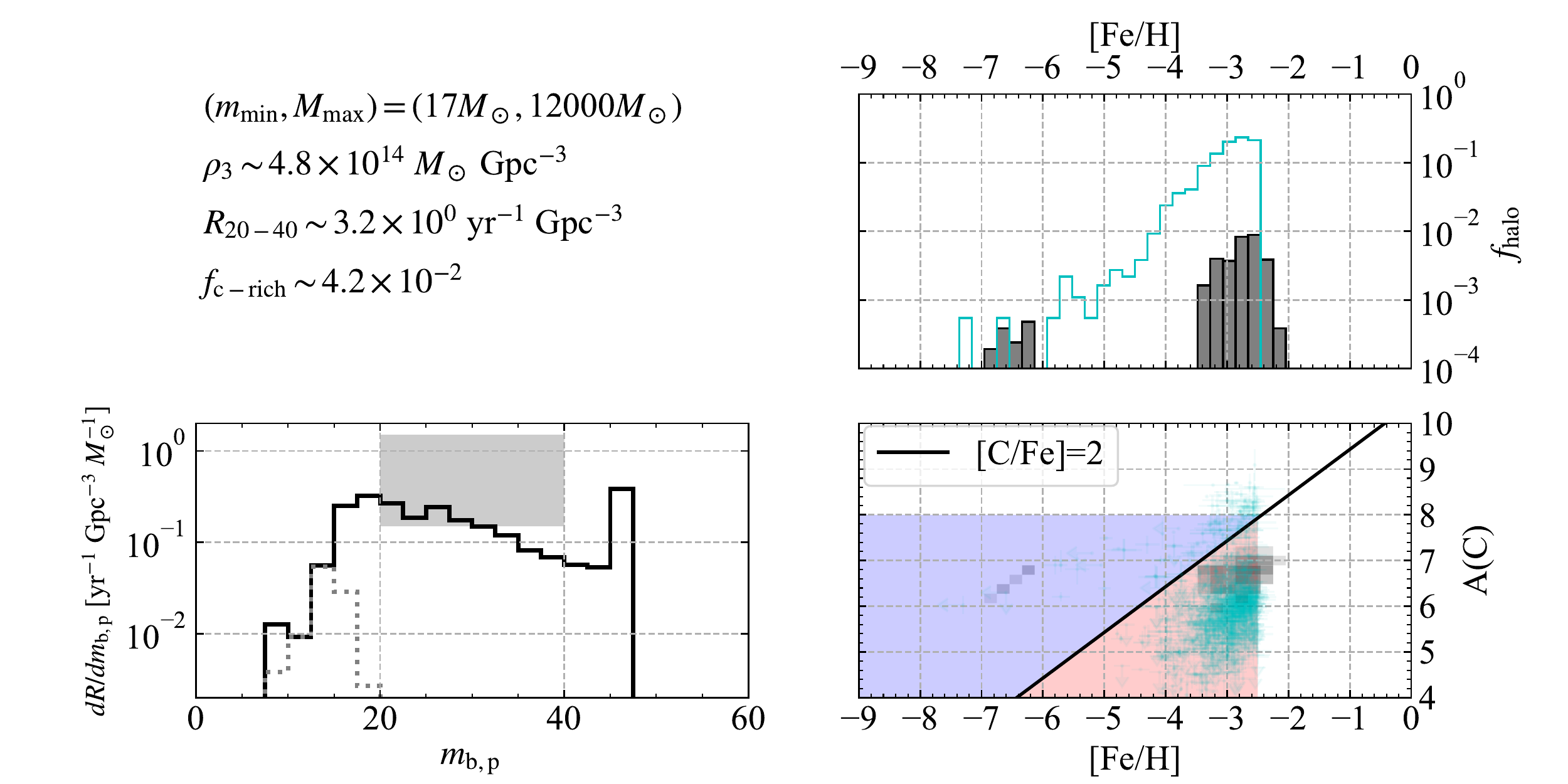} 
 \end{center}
\caption{The same as Figure \ref{fig:figExample1} except for input
  parameters of $(\msmin, \mlmax)=(17\msun,12000\msun)$. In the bottom
  left panel, the lower-mass peak is formed through the same mechanism
  as in Figure~\ref{fig:figExample1}. The higher-mass peak in
  Figure~\ref{fig:figExample1} disappears in this figure, since the
  higher-mass peak needs many binary stars with $\lesssim 17$ $\msun$
  secondary stars.} \label{fig:figExample3}
\end{figure*}

In order to increase $\fcemp$, we adopt $(\msmin, \mlmax) = (17\msun,
12000\msun)$ as seen in Figure \ref{fig:figExample3}. In this case,
the third necessary condition are satisfied as well as the first and
second conditions. Since we increase $\msmin$, minihalos have Pop III
stars with $\mzams \lesssim 20 \msun$ and CCSNe at a small
probability. Then, a significant fraction of minihalos are not
polluted by iron. There are no EMP stars with ${\rm [Fe/H]} \sim
-6$--$-3$ in our model in contrast to the observation of EMP
stars. However, this may be reconciled when we take into account
inhomogeneous mixing of Pop III supernova ejecta in minihalos.

In order to assess the Pop III BH merger scenario, previous studies
have focused on the secondary necessary condition. Thus, they have
only pointed out that minihalos should form a larger amount of Pop III
stars than predicted theoretically by a factor of more than 10. In
addition to that, we find a new constraint on $\msmin$: $15 \lesssim
\msmin/\msun \lesssim 27$ by taking into account the third necessary
condition. The allowed range $\mlmax$ becomes narrower when $\msmin$
deviates from $\sim 20$ $\msun$. In other words, $\msmin$ should be
close to the critical mass where the fate of a Pop III star transits
from a CCSN to a FSN. The critical mass is $\sim 20$ $\msun$ in our
adopted supernova model. Otherwise, minihalos can form only one type
of carbon-rich and carbon-normal EMP stars.

\section{Summary and discussion}
\label{sec:SummaryAndDiscussion}

We examine the Pop III BH merger scenario under the hypothesis that
EMP stars consist of primordial gas and Pop III supernova ejecta. We
suppose that the scenario is valid if the following three necessary
conditions are satisfied. First, Pop III formation rate is $\dthree
\le 10^{15}$ $\msun$ $\cgpc$. Second, Pop III BHs merge at a rate of
$3 \le \ratetarget/(\pyr \cgpc) \le 30$. Third, EMP stars have the
number ratio $0.01 \le \fcemp \le 1$. In order to obtain Pop III BHs
and EMP stars, we construct a simple formation model of Pop III stars,
and simulate the evolution of Pop III stars by binary population
synthesis technique.

We find that there is a region satisfying the three necessary
conditions. The Pop III formation rate should be $\dthree \gtrsim 3
\times 10^{14}$ $\msun$ $\cgpc$. This means that each minihalo should
form Pop III stars with $\mlmax \sim 10^4$ $\msun$, or $10^3 - 10^4$
$\msun$ in total. Although this formation rate does not violate the
first necessary condition, it is much larger than numerically
predicted by \citet{2020MNRAS.492.4386S}. Similar things have been
already pointed out by \citet{2016MNRAS.460L..74H}.

We newly find that the minimum mass of Pop III stars should be $15
\lesssim \msmin/\msun \lesssim 27$. If $\msmin/\msun \lesssim 15$ or
$\msmin/\msun \gtrsim 27$, $\fcemp$ is smaller or larger than the
third necessary condition, respectively. This is because each minihalo
gets too much CCSNe and FSNe for $\msmin/\msun \lesssim 15$ and
$\msmin/\msun \gtrsim 27$, respectively. Moreover, the allowed region
becomes narrower with $\msmin$ deviating from $\sim 20$ $\msun$, which
is the boundary mass between CCSNe and FSNe.

We compare the new constraint with $\msmin$ obtained from numerical
simulations. \citet{2014ApJ...792...32S} have shown that the number of
Pop III stars is sharply decreased just below $\sim 10$ $\msun$. On
the other hand, in \citet{2014ApJ...781...60H} (see also
\cite{2015MNRAS.448..568H}), the number of $10$ $\msun$ Pop III stars
is much smaller than that of $20$ $\msun$, which can be interpreted as
$10 \lesssim \msmin/\msun \lesssim 20$. From these numerical
simulations, we can regard $10 \lesssim \msmin/\msun \lesssim 20$,
however cannot strictly determine $\msmin$. Thus, we do not exclude
possibility of the Pop III BH merger scenario, under our assumption
that EMP stars are formed from primordial gas mixed with Pop III
supernova ejecta.

We have to note that recent studies have also found the formation of
low-mass Pop III stars \citep{2008ApJ...677..813M,
  2011ApJ...727..110C, 2011Sci...331.1040C, 2011ApJ...737...75G,
  2012MNRAS.424..399G, 2013MNRAS.435.3283M, 2014ApJ...792...32S,
  2016MNRAS.463.2781C}. Such low-mass Pop III stars can have $\lesssim
1$ $\msun$. However, the number of such low-mass Pop III stars is much
smaller than Pop III stars with $\gtrsim 10$ $\msun$. The presence of
such low-mass Pop III stars can be negligible for our results.

We emphasize that the new constraint on $\msmin$ ($15 \lesssim
\msmin/\msun \lesssim 27$) is necessary only when we claim the Pop III
BH merger scenario. Note that Pop III stars are a major origin of
observed BH mergers with more than 20 $\msun$ in the Pop III BH merger
scenario. In other words, Pop III stars can have $\lesssim 15$ $\msun$
stars (say $10$ $\msun$ stars or less massive stars) if we do not
adopt the Pop III merger scenario. The new constraint on $\msmin$ can
be applicable when we reconcile the Pop III BH merger scenario with
$0.01 \le \fcemp \le 1$.

Hereafter, we make caveats on our model. We treat $\mlmax$ as a free
parameter, while we fix $\nhalo$ as $\sim 10^{11}$ $\cgpc$. However,
$\mlmax$ and $\nhalo$ may be
anti-correlated. \citet{2020MNRAS.492.4386S} have performed numerical
simulations to follow cosmic chemical evolution as well as Pop III
star formation. Since we assume that more Pop III stars are formed
than \citet{2020MNRAS.492.4386S} expected, Pop III's surrounding gas
should be polluted more rapidly. The pollution should stop forming Pop
III stars earlier, and start forming Pop I/II stars earlier. Thus, if
we assume that each minihalo yields a larger amount of Pop III stars
than their results (i.e. $\mlmax \gg 600$ $\msun$), we should decrease
the number density of minihalos (i.e. $\nhalo \ll 10^{11}$
$\cgpc$). In that case, the Pop III BH merger scenario should be
difficult to be valid. In this paper, we do not take into account
this, because we simply control Pop III star formation rate.

We prepare the three necessary conditions as described in section
\ref{sec:NecessaryConditions}. It might be possible to add other
necessary conditions. As for merging BHs, GW observations obtain not
only BH masses but also BH spins. We might account for BH spins for
necessary conditions. However, we do not do so. This is because the
estimates of BH spins are less reliable and less constrained than BH
masses \citep{2021arXiv211103606T}. Thus, we adopt only BH masses for
the necessary conditions. As for EMP stars, chemical elements other
than carbon are also observed. Nevertheless, we do not take into
account them. This is because carbon abundance is the most
characteristic and most observed elements in EMP stars, and is the
most sensitive to whether CCSNe or FSNe happen.

In our supernova model, Pop III stars with $\mzams/\msun \lesssim 20$
and $\gtrsim 20$ deterministically cause CCSNe and FSNe, respectively,
except that binary interactions slightly change their fates. On the
other hand, recent supernova studies have suggested that stars with
$\mzams = 10-30$ $\msun$ cause CCSNe stochastically
\citep{2012ApJ...757...69U, 2015ApJ...801...90P, 2016ApJ...821...38S,
  2016MNRAS.460..742M, 2018ApJ...860...93S, 2021ApJ...909..169K}. This
may largely affect our results. However, we do not adopt such
supernova models. This is because these models have not been prepared
to use in binary population synthesis technique. These models tell us
whether stars experience CCSNe or direct collapses to BHs, but do not
include FSNe which is needed in binary population synthesis technique.

\begin{ack}
  We thank organizers of the first star and first galaxy conference
  2020 in Japan for giving us a good opportunity to start our
  collaboration. This study was supported in part by Grants-in-Aid for
  Scientific Research (19K03907, 20H00174, 20H01904, 21K13915) from the Japan
  Society for the Promotion of Science and Grants-in-Aid for
  Scientific Research on Innovative areas (17H06360, 18H05437,
  20H04747) from the Ministry of Education, Culture, Sports, Science
  and Technology (MEXT), Japan.
\end{ack}


\end{document}